\documentclass[%
 reprint,%<- UN-COMMENT THIS LINE
superscriptaddress,%<- UN-COMMENT THIS LINE
 amsmath,amssymb,
aps, %<- UN-COMMENT THIS LINE
prl, %<- UN-COMMENT THIS LINE
showkeys,
]{revtex4-2}

\usepackage{graphicx}% Include figure files
\usepackage{dcolumn}% Align table columns on decimal point
\usepackage{bm}% bold math
\begin{document}

\preprint{APS/123-QED}

\title{Characterization of  Quasi-Keplerian, Differentially Rotating, Free-Boundary Laboratory Plasmas}

\author{V. Valenzuela-Villaseca}
\email{vicente.valenzuela-villaseca17@imperial.ac.uk.}
\author{L. G. Suttle}
\author{F. Suzuki-Vidal}
\author{J. W. D. Halliday}
\author{S. Merlini}
\author{D. R. Russell}
\author{E. R. Tubman}
\altaffiliation[Current address: ]{Lawrence Livermore National Laboratory, 7000 East Ave, Livermore, California 94550, USA.}
\author{J. D. Hare}
\altaffiliation[Current address: ]{Plasma Science and Fusion Center, Massachusetts Institute of Technology, Cambridge, Massachusetts 02139, USA. }
\author{J. P. Chittenden}
\affiliation{Blackett Laboratory, Imperial College, London SW7 2BW, UK.}
\author{M. E. Koepke}
\affiliation{Department of Physics, West Virginia University, USA.}
\author{E. G. Blackman}
\affiliation{Department of Physics and Astronomy, University of Rochester, USA.}
\author{S. V. Lebedev}
\affiliation{Blackett Laboratory, Imperial College, London SW7 2BW, UK.}

\date{\today}

\begin{abstract}
We present results from pulsed-power driven differentially rotating plasma experiments designed to simulate physics relevant to astrophysical disks and jets. In these experiments, angular momentum is injected by the ram pressure of the ablation flows from a wire array Z pinch. In contrast to previous liquid metal and plasma experiments, rotation is not driven by boundary forces. Axial pressure gradients launch a rotating plasma jet upwards, which is confined by a combination of ram, thermal, and magnetic pressure of a surrounding plasma halo. The jet has subsonic rotation, with a maximum rotation velocity $23 \pm 3$ km/s. The rotational velocity profile is quasi-Keplerian with a positive Rayleigh discriminant $\kappa^2 \propto r^{-2.8\pm0.8}$ rad$^2$/s$^2$. The plasma completes $0.5 - 2$ full rotations in the experimental time frame ($\sim 150$ ns).
%\begin{description}
%use the optional argument of the \verb+\item+ command to give the category of each %item. 
%\end{description}
\end{abstract}

%Use showkeys class option if keyword
\maketitle

%\tableofcontents

Differentially rotating magnetohydrodynamical (MHD) accretion flows orbiting a central object are ubiquitous in the universe \cite{J.FrankA.King2002}. The observational evidence of accretion towards the central object requires angular momentum transport to be far more efficient than can be provided by kinematic viscosity. The enhancement is often modelled as anomalous turbulent viscosity \cite{Shakura1973}. However, the Rayleigh criterion guarantees the hydrodynamical stability of Keplerian flows they have a specific angular momentum distribution $\ell(r)$  increasing with radius as $\ell = \Omega r^2 \propto r^{1/2}$ ($r$ is the radial position and $\Omega$ the rotation velocity) \cite{Chandrasekhar1961}. Luckily,  a  magnetized flow with the same rotation profile is unstable to the Magneto-Rotational Instability (MRI) \cite{Velikhov1959,Balbus1991,Goodman1994}, which is triggered when a differentially rotating flow has an angular frequency distribution $\Omega(r)$ that monotonically decreases with radius. Indeed, Keplerian flows satisfy $\Omega \propto r^{-3/2}$.

Existing laboratory experiments use the Taylor-Couette geometry to realise steady-state rotating MHD flows and study instabilities, turbulence and angular momentum transport developing gradually over hundreds of rotation periods. These experiments control the rotation profile from the edges of the flow, either by spinning the vessel containing a liquid (such as water or aqueous-glycerol \cite{Ji2006}, or sodium or gallium alloys under an external axial or helical magnetic field \cite{Ji2001,Goodman2002,Hollerbach2005,Stefani2006,Liu2006}), or by applying electrical currents from the edge of Hall plasmas confined by permanent magnets at the boundary \cite{Collins2012,Flanagan2020,Milhone2021}.

In this Letter, we present results from the Rotating Plasma Experiment (RPX), a novel pulsed-power driven platform to study free-boundary differentially rotating plasmas. The flow is quasi-Keplerian (qK), i.e. $d\Omega/dr < 0$ and $d\ell/dr >0$, and allows the development of effects with growth times comparable to the rotation period (e.g. the MRI \cite{Balbus1991}). The measured Rayleigh discriminant \cite{Balbus2012} is $\kappa^2 \propto r^{-2.8\pm0.8}$ rad$^2$/s$^2$ $>0$. Thus, the flow rotation profile meets the conditions to generate the MRI in a collisional plasma. Additionally, the free-boundary design allows the formation of axial plasma jets launched from the ends of the plasma column, without the flow perturbations characteristic of rigid wall containment \cite{Gissinger2011,Edlund2015,Balbus2017,Caspary2018}. The axial jet rotates at a maximum velocity $\gtrsim 20$ km/s, undergoing between $0.5$ and $2$ full rotations for the duration of the experiment.

\begin{figure}
    \centering
    \includegraphics[width=8.6cm]{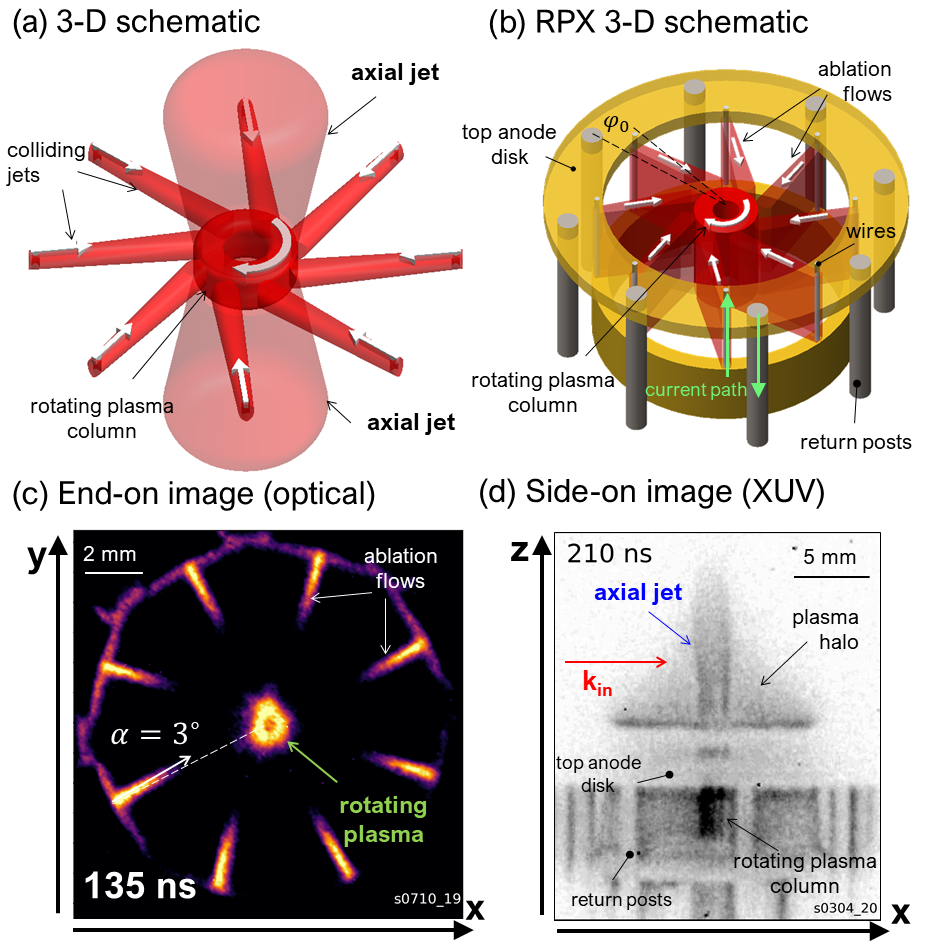}
    \caption{Schematic diagrams of experimental set up and self-emission images.(a) 3-D schematic of inertially driven rotating plasmas. (b) 3-D schematic of experimental hardware.  Axial jets are not shown. (c) End-on optical self-emission image.(d) Extreme ultra-violet (XUV) image of rotating plasma.}
    \label{fig:schem_self-emission}
\end{figure}

The experimental platform is based on inertially driving a rotating plasma by the oblique collision of multiple plasma flows (colliding jets), which continuously inject both mass and angular momentum to a rotating plasma column, as shown in FIG. \ref{fig:schem_self-emission}a. The colliding jets allow radial confinement of the column via ram pressure gradient providing centripetal acceleration \cite{Ryutov2011}. At the same time, axial pressure gradients launch rotating axial jets from the top and bottom ends of the rotating plasma column (FIG. \ref{fig:schem_self-emission}a). On RPX, the colliding jets emerge from the ablation of a cylindrical arrangement of thin metallic wires driven by a pulsed-power generator. These plasma jets are termed ablation flows \cite{Bocchi2013-ApJ,Bocchi2013-HEDP,Bennett2015}, as shown in FIG. \ref{fig:schem_self-emission}b. They have densities of a few $10^{18}$ cm$^{-3}$, and propagate with velocities $4\times 10^4 - 1\times 10^5$ m/s ($M > 5$) and characteristic temperature $T\sim 10$ eV \cite{Harvey-Thompson2012-PRL,Harvey-Thompson2012-PoP,Swadling2013}. The inwards, off-axis trajectory of the ablation flows is set by the azimuthal and radial components of magnetic field of the applied current. The azimuthal component is generated by the current passing through the wires, whereas the radial component is introduced by return posts slightly off-set relative to the wires by an angle $\varphi_0$ (FIG. \ref{fig:schem_self-emission}b). The duration of the experiment is limited by the initial mass of the wires driven by the pulsed-power generator \cite{Lebedev2001}. In the experiments presented on this paper, rotation driven for up to $270$ after current start.

The experiments were conducted on the MAGPIE pulsed-power generator which delivers $1.4$ MA peak electrical current with $240$ ns rise-time \cite{Mitchell1996}. The plasma source is an ablating aluminium wire array (eight 40 $\mu$m wires, $16$ mm diameter, $10$ mm high). The wire array is surrounded by eight 1 mm diameter stainless steel return posts, offset by $\varphi_0 = 13^{\circ}$ and positioned on a $22$ mm diameter, as shown in FIG. \ref{fig:schem_self-emission}b. In each experiment, measurements of the temporal evolution of the plasma, density, temperature, and velocity are obtained by using a multi-diagnostic suite consisting of self-emission (optical and XUV, 5 ns time resolution) and laser probing (interferometry and optical Thomson Scattering [TS]) \cite{Swadling2013,Swadling2014,Suttle2021}.

FIG. \ref{fig:schem_self-emission}c shows one of the end-on optical self-emission images (5 ns time resolution, $>600$ wavelength nm) of the ablation flows and the formation of the rotating plasma. The ablation flows propagate inwards, with an off-set propagation angle $\alpha=3^{\circ}\pm 1^{\circ}$ (FIG. \ref{fig:schem_self-emission}c). As the ablation flows propagate inwards, their emission reduces due to radiative cooling \cite{Lebedev2005}. Considering the ablation velocity $V_{ab}=6\times 10^{4}$ m/s \cite{Harvey-Thompson2012-PRL} and array radius $r=8\times10^{-3}$ m, the specific angular momentum introduced by each wire ablation is $\ell = rV_{ab}\sin(\alpha) =25\pm 10$ m$^2$/s. The $12$ optical self-emission images obtained from each experiment indicate that $\alpha$ can slightly vary by $\sim 1^{\circ}$ between ablation flows in the same experiment due to uneven current distribution through the load.

The formation of the rotating plasma is shown in FIG. \ref{fig:schem_self-emission}c, observed as a bright ring with a reduced intensity on-axis. This structure is consistent a dense plasma shell with a density depletion on-axis sustained by a centrifugal barrier. Thus, the plasma accumulates where the centrifugal force is balanced by the inwards ram pressure of the ablation flows. This hollow structure is sustained from $130$ ns (formation time) to $210$ ns.

Above the array, axial plasma outflows are observed by a pair of 4-frame, XUV cameras ($100$ $\mu$m pinhole, $1$ $\mu$m Mylar filter, $>40$ eV photon energy, 5 ns time resolution), as shown in FIG. \ref{fig:schem_self-emission}d. A highly collimated jet is launched by axial thermal and magnetic pressure gradients from the formed rotating plasma column. By tracking the jet length across different frames, the estimated axial velocity is $u_z=100 \pm 20$ km/s. The jet has a divergence angle of $3^{\circ} \pm 1^{\circ}$ and there is no visible development of typical MHD instabilities of Z pinch plasmas \cite{Bott2006,Veloso2015}, prior to $> 270$ ns. The axial jet is surrounded by a plasma halo, indicated in FIG. \ref{fig:schem_self-emission}d.

\begin{figure}
    \centering
    \includegraphics[width=8.6cm]{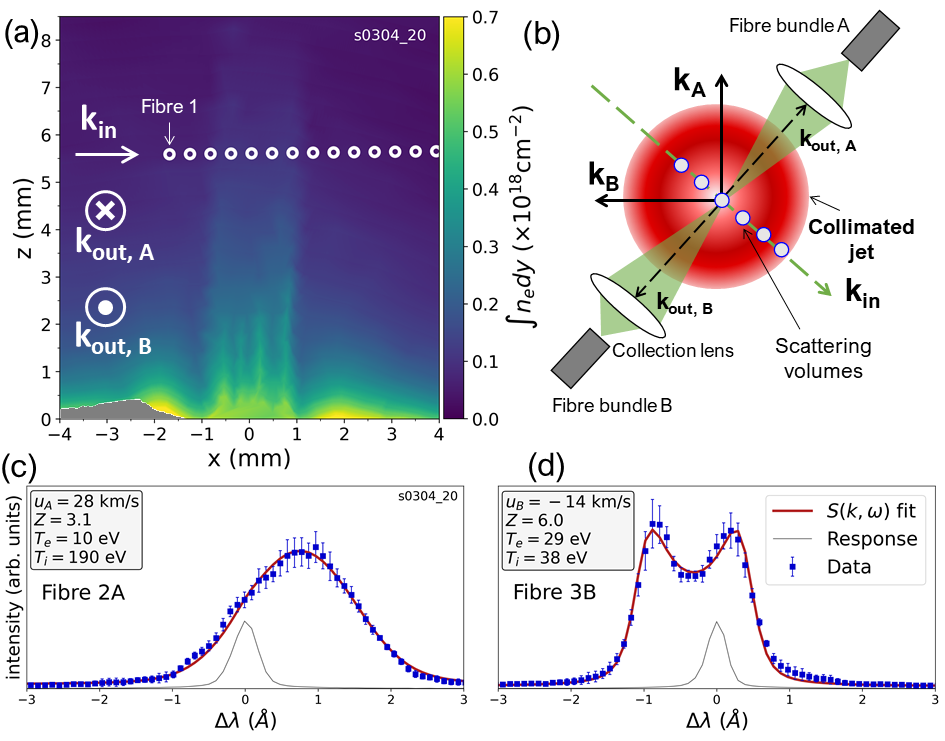}
    \caption{(a) Line integrated electron density map. TS volumes are overlaid in white circles. Additional arrows indicate the scattering geometry. (b) Schematic TS set-up and vector diagram. (c) TS spectrum from outside the jet (plasma halo). Best fit is shown in red. The Spectrometer response is shown in grey. Fitting parameters shown in grey box. (d) TS spectrum from inside the jet. The characteristic double-peak of the ion-acoustic wave feature is observed. }
    \label{fig:side-on_interferometry}
\end{figure}

Measurements of the outflows electron density were obtained using a Mach-Zehnder interferometer ($532$ nm wavelength, $0.5$ ns FWHM) are presented in FIG. \ref{fig:schem_self-emission}d, with the probing beam passing side-on. The raw inteferogram was analysed using the MAGIC2 code \cite{Swadling2013,Hare2019} to construct a line integrated electron density $\int n_e dy$ map of the outflows, presented in FIG. \ref{fig:side-on_interferometry}a. Our coordinate system is such that the height $z=0$ coincides with the upper surface of the top anode disk.

\begin{figure}
    \centering
    \includegraphics[width=8.6cm]{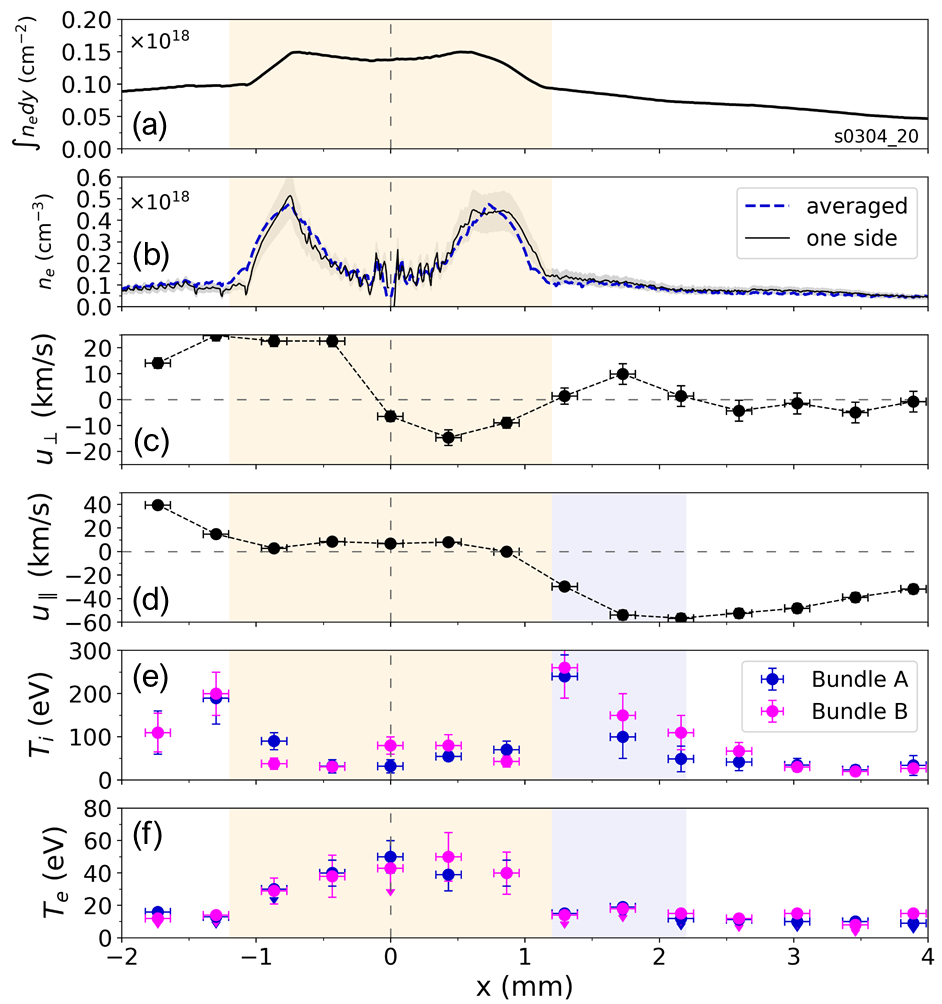}
    \caption{Plasma parameters. (a) Line-integrated electron density lineout at the same height as the scattering volumes. Uncertainty was estimated to be $20\%$. (b) Abel inverted electron density. Left- and right-hand-side inversions (black lines) and averaged inversion (dashed blue line). (c) Rotation velocity. (d) Radial velocity. (e) Ion temperature. (d) Electron temperature. Downwards pointing arrows indicate the value is an upper constraint.}
    \label{fig:TS-rotation}
\end{figure}

A characterization of the density profile, velocity components in the plane of rotation, and ion and electron temperature of the outflows is presented in FIG. \ref{fig:TS-rotation}. 
A line integrated electron density lineout at the height of the TS measurements (see FIG. \ref{fig:side-on_interferometry}a) is presented in FIG. \ref{fig:TS-rotation}a. The lineout was Abel inverted using a onion-peeling method \cite{Dasch1992} independently on each side of the rotation axis, and then the two sides are averaged, to obtain an electron number density. The density profile of the plasma in this region consists of an outer plasma halo at $\sim 0.2 \times 10^{18}$ cm$^{-3}$ surrounding a hollow axial jet of maximum density $\sim 0.5 \times 10^{18}$ cm$^{-3}$ on the shell and axial density comparable to the halo.

Local measurements of velocity in the plane of rotation and temperature are obtained using the TS diagnostic. Scattered spectra are collected from $14$, $200$ $\mu$m diameter scattering volumes located across the collimated jet and the surrounding plasma halo, as shown in FIG\ref{fig:side-on_interferometry}a. The scattered light is collected by two linear arrays of $100$ $\mu$m diameter optic fibres (labelled A and B), located at $\pm 90^{\circ}$ relative to the probing wave-vector $k_{in}$, as shown in FIG\ref{fig:side-on_interferometry}b. Fits of the TS spectra (FIG \ref{fig:side-on_interferometry}c and d) yield measurements of the velocity components along the scattering vectors $\mathbf{k_A}\equiv\mathbf{k_{out, A}}-\mathbf{k_{in}}$ and $\mathbf{k_B}\equiv\mathbf{k_{out, B}}-\mathbf{k_{in}}$, and the temperature $T_i$ and the product of the average charge state and electron temperature $ZT_e$. This product is decoupled using the non-LTE code SpK \cite{Niasse2011,Hare2017-Thesis}. The fit uses the electron density obtained from interferometry shown in FIG. \ref{fig:TS-rotation}b. From the velocity components $u_{A,B} \equiv \mathbf{u}\cdot\mathbf{k_{A,B}}$, the new components, along and perpendicular to $\mathbf{k_{in}}$ are calculated as

\begin{equation}
    u_{\parallel} = \frac{1}{\sqrt{2}}(u_A+u_B), u_{\perp} = \frac{1}{\sqrt{2}}(u_A-u_B). 
\end{equation}

The laser beam passes approximately through the rotation axis with Fiber $5$ viewing the position $x=0$ with an estimated accuracy of $200$ $\mu$m, however we cannot exclude an off-set of the jet in the perpendicular, $y-$direction (out-of plane in FIG. \ref{fig:side-on_interferometry}a). If there is no off-set of the probe beam along $y$, the absolute values of the components $(u_{\parallel},u_{\perp})$ correspond exactly to the radial and rotation velocity components $(u_r, u_{\theta})$. We show below that the interpretation of the data is not sensitive to the off-set of the beam relative to the axial jet along the $y-$axis.

The rotation velocity distribution is shown in FIG. \ref{fig:TS-rotation}c. Inside the jet, rotation is manifested as the change in sign in the velocity about the axis (orange region in FIG \ref{fig:TS-rotation}c) exhibiting a maximum velocity of $23\pm 3$ km/s. 
Flow velocity distribution is not symmetric about the jet axis. 
This could be imprinted from the discrete nature of the driver, and/or due to slight misalignment in hardware construction. FIG \ref{fig:TS-rotation}d shows the radial component of velocity. The plasma from the halo propagates inwards at a maximum velocity of $45 \pm 5$ km/s, before decelerating in the vicinity of the jet. 
We call this the deceleration region (blue region in FIG \ref{fig:TS-rotation}d). 
This component reaches a minimum inside of the jet, pointing along $\mathbf{k_{in}}$, which indicates that probe beam is misaligned with the rotation axis.

Ion and electron temperature distributions obtained by fitting the spectra from each fibre bundle independently are shown in FIG. \ref{fig:TS-rotation}e. 
At large radii ($x>3$ mm) $T_i \approx T_e \sim 10$ eV and gradually increase as the plasma approaches the deceleration region. However, at the deceleration region, the ion temperature of the inflowing plasma increases by a factor of $5$ over a length $\sim 1$ mm, reaching $T_i\sim 250$ eV, whereas the electrons increase to $T_e \lesssim 20$ eV. 
It is inferred that the inflow undergoes a transonic deceleration: from a sonic Mach number $M_s\sim 3$ at $x=2.5$ mm, to $M_s\lesssim 1$ in the deceleration region. The electron-ion equilibration time in the deceleration region $\bar{\nu}_{e\setminus i}^{-1} = 20 \pm 5$ ns \cite{Huba2016}, equal to the transit time in this region. 

Inside the jet, $T_i$ range between $\sim 50$ eV and $100$ eV, whereas $T_e$ range between $\sim 30$ eV and $50$ eV, exhibiting a gradual increase towards the axis. This increase $T_e$ is consistent with efficient electron-ion thermal equilibration ($\bar{\nu}_{e\setminus i}^{-1} = 18 \pm 5$ ns). However, radiative cooling prevents full equilibration as the radiative cooling time is $\tau_{cool} \sim 10$ ns \cite{Suzuki-Vidal2015,Russell2021} ($\approx \bar{\nu}_{e\setminus i}^{-1}$), keeping the electrons thermally uncoupled to the ions. The inferred rotational Mach number $M_s \sim 0.5$, meaning that rotation is subsonic.

\begin{figure}
    \centering
    \includegraphics[width=8.6cm]{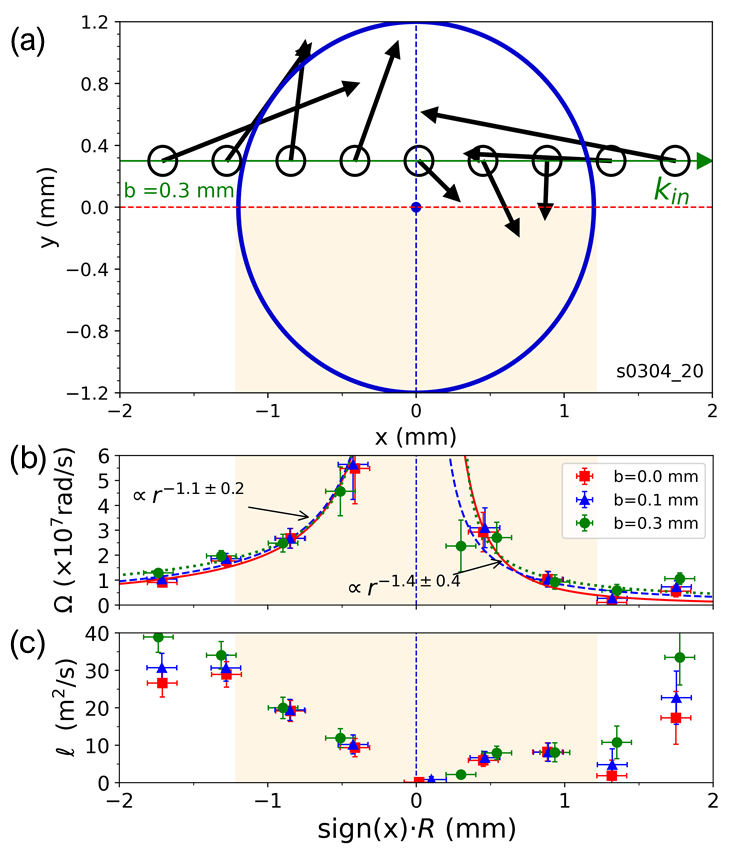}
    \caption{Rotation velocity distribution depending on TS beam impact parameter $b$. (a) Diagram of the velocity map (black arrows) of each TS volume (black circles). Example impact parameter $b=0.3$ mm is shown. (b) Angular frequency distribution. (c) Angular momentum distribution.}
    \label{fig:TS-rotmap}
\end{figure}

Further characterization of the velocity distribution is presented. FIG. \ref{fig:TS-rotmap}a shows an end-on 2-D velocity map constructed from the TS measurements, where the positions of the scattering volumes are translated according to an impact parameter $b$, defined as the orthogonal distance from the probing beam to the rotation axis (i.e. along $y$). The scattering volumes are located at positions $x_i$ along the $x$-axis, thus the angular frequency at radius $R_i = \sqrt{x_i^2+b^2}$ is given by

\begin{equation}
    \Omega(R_i) = \frac{x_iu_{\perp}-bu_{\parallel}}{R_{i}^2}.
\end{equation}

FIG. \ref{fig:TS-rotmap}b, shows the calculated angular frequency distributions for three values of the impact parameter: $b=0$ mm (i.e. beam passing through the rotation axis), $b=0.1$ mm, and $b=0.3$ mm. The result shows the distribution does not significantly change with the impact parameter. 
The angular frequency distribution for each value of $b$ is fitted independently either side of the axis using a power-law $\propto r^{\gamma}$. For the case $b=0$, the point  closest to the axis was not considered. The variation of values of $b$ was used to find a range of values of $\gamma$ indicated in the Figure. The Rayleigh discriminant can be calculated as $\kappa^2 \equiv r^{-3}d(r^4\Omega^2)/dr \propto (4+2\gamma)r^{2\gamma} = r^{-2.8\pm0.8} > 0$. The physical meaning of this value of $\kappa^2$ is twofold. Firstly, its positive value implies that the flow is hydrodynamically stable against axisymmetric perturbations. This is because a linearly perturbed trajectory would oscillate around an unperturbed one (having same initial conditions) with frequency $\kappa$, termed epicyclic frequency \cite{Balbus1998}. Non axisymmetric modes are unstable when $\ell(r) =$ constant, which is not satisfied by our plasma (see FIG. \ref{fig:TS-rotmap}c) \cite{Papaloizou1984,Balbus1998}. Secondly, a magnetized, rotating flow can be MRI unstable when $d\Omega/dr<0$. It is explicit from the definition of $\kappa^2$ that both hydro-stability and magneto-rotational instability are satisfied simultaneously when $-2<\gamma<0$. This case is called quasi-Keplerian rotation. In that sense, $\kappa^2$ fully characterizes the differential rotation profile.

Additional conditions are required for the MRI to develop. The standard MRI requires a vertical field as an initial condition \cite{Balbus1991}, and even the Helical MRI ( which includes an azimuthal magnetic field) is suppressed when the field pitch angle $\theta = \tan^{-1}(B_z/B_{\theta})\rightarrow 0$ \cite{Liu2006}. In that regard, a vertical magnetic field (absent in these experiments) is essential for the instability. However, this paper demonstrates that this free-boundary laboratory platform is able to produce rotating flows with the correct profile to generate the MRI in a laboratory plasma. The characteristic length-scales for both magnetic diffusion ($\ell_{\eta}\sim 70$ $\mu$m) and viscous diffusion ($\ell_{\nu}\sim 0.5$ $\mu$m), where $\ell_{\eta}$ and $\ell_{\nu}$ are such that the magnetic and fluid Reynolds numbers are of order unity \cite{Ryutov1999,Huba2016}, meaning that dissipation is negligible. Thus, the MRI fastest growin mode would develop on timescales $|\omega_{max}|^{-1}=|\kappa^2/4\Omega - \Omega|^{-1}=150\pm100$ ns, comparable to the orbital period \cite{Balbus1998}. It is interesting that the observed stability indicates that any perturbations introduced by the ablation into the rotating plasma are insignificant for the evolution of the rotating axial jet.

In summary, we have presented an experimental characterization of free-boundary rotating laboratory plasmas which launch axial jets with a hollow density structure. The measured rotation velocity profile corresponds to quasi-Keplerian rotation with Rayleigh discriminant $\kappa^2 \propto r^{-2.8\pm0.8} > 0$, indicating that the flow is hydrodynamically stable but potentially MHD unstable. In the duration for the experiment, the plasma completes $0.5 - 2$ full rotations (depending on radial position). This is limited to the mass per unit length of each aluminium wire, but it can be increased by using wires of greater diameter. Although current pulsed-power experiments are yet too short-lived to reach the MRI nonlinear regime and associated turbulent angular momentum transport, these experiments provide an essential step in establishing a laboratory astrophysics platform for rotating plasma flows and motivate further developments of diagnostics. Even unambiguous identification the linear amplification of an initial seed magnetic field growth by differential rotation in this platform, let alone the fully developed MRI, would be a novel proof-of-principle measurement. These results are a first step towards new efforts to generate and study the MRI, dynamo effects and collimation of jets in a single plasma experiment, where effects such as radiative cooling are also important. Future experiments will study the magnetic field evolution to investigate the existence shear-flow magnetic dynamo on RPX.

%%%%%%%%%%%%%%%%
%%%%%%%%%%%%%%%%
%%%%%%%%%%%%%%%%

\begin{acknowledgments}
This work was supported in part by NNSA under DOE Cooperative Agreement No DE-SC0020434 and DE-NA0003764, and the Royal Astronomical Society. Vicente Valenzuela-Villaseca is funded by the Imperial College President's PhD Scholarships. We are greatly thankful to the Mechanical Workshop at the Imperial College Department of Physics, and in particular to Mr David Williams, for their contribution to hardware manufacture. We are thankful to Thomas Varnish for writing some of the codes used in the analysis.
\end{acknowledgments}

%\nocite{*}

\bibliography{bibliography_apssamp}% Produces the bibliography via BibTeX.

\end{document}